\documentclass[twoside,slac_one]{revtex4}
\usepackage{graphicx}
\usepackage{fancyhdr}
\usepackage{amsmath} 
\usepackage{bm}
\usepackage{amsxtra}
\usepackage{amssymb}
\usepackage{amsthm}
\usepackage{latexsym}
\usepackage{lscape}

\begin{document}
\title{Measurement of elliptic and higher order flow harmonics at $\sqrt{s_{NN}}=2.76$ TeV Pb+Pb collisions with the ATLAS Detector}


\author{Soumya Mohapatra for the ATLAS Collaboration}
\affiliation{Stony Brook University, NY 11794, USA}

\begin{abstract}
The measurements of flow harmonics $v_2$-$v_6$ using the event plane and two particle correlations methods in broad $p_T$, $\eta$ and centrality ranges using the ATLAS detector at LHC are presented. ATLAS recorded about 9 $\mu \textrm{b}^{-1}$ of lead-lead collision data in the 2010 heavy ion run. The full azimuthal acceptance of the ATLAS detector in $\pm2.5$ units of pseudorapidity for charged hadrons and the large amount of data allows for a detailed study of the flow harmonics. The $p_T$, centrality and $\eta$ ranges where the two methods give consistent $v_n$ and where they disagree are discussed.  It is shown that the ridge as well as the so called ``mach-cone" seen in two particle correlations are largely accounted for by the collective flow. Some scaling relations in the $p_T$ dependence of the $v_n$ are also discussed.
\end{abstract}

\maketitle

\thispagestyle{fancy}


\section{Introduction}

A central goal of the heavy ion program at the LHC is to understand the properties of the hot and dense matter produced in relativistic heavy ion collisions, commonly termed as the quark gluon plasma. One of the ways to study its properties is to perform high accuracy measurements of the azimuthal yields of the produced particles. Due to anisotropic pressure gradients, the ``fireball" produced in the collisions expands differently in different directions with more particles emitted along the larger gradients. This anisotropy in the particle yields can be expressed as a Fourier series:

\begin{equation}
\frac{dN}{d\phi} \propto 1+2\sum_{n=1}^{\infty} v_n^a \cos(n(\phi-\Psi_n))
\label{eq_flow}
\end{equation}


Where $\phi$ is the azimuthal angle of the produced particles, $v_n$ is the magnitude of the $n^{th}$ order harmonic flow and $\Psi_n$ is the orientation of the corresponding harmonic plane \cite{bib_q_vector}. The superscript `$a$' is used to indicate particle species, $p_T$ bin, centrality, etc. 

The elliptic flow $v_2$ has been studied extensively as it usually dominates other coefficients, being strongly influenced by the elliptic geometry of the overlap region between the colliding nuclei. However, the initial geometry of the produced fireball may have angular moments of several orders: elliptic, triangular, etc due to fluctuations in the spatial positions of the nucleons in the colliding nuclei. These higher order moments in the initial geometry can give rise to measurable higher order flow ($n>3$). 

A better understanding of the higher order $v_n$ can explain the origin of the ridge, an elongated structure along $\Delta\eta$ at $\Delta\phi\sim 0$ \cite{bib_ridge_cone1} and the so-called ``mach-cone", a double hump structure on the away side \cite{bib_ridge_cone2} seen in two particle correlations. These were initially interpreted as response of the medium to the energy deposited by the quenched jets. However, recent studies \cite{bib_v3cone_} have shown that higher order flow harmonics make a significant contribution to these structures. The $v_n$ can also be used to obtain information about the initial geometry and transport properties of the medium, for instance the viscosity to entropy ratio: $\eta/s$ \cite{bib_visc_calc1,bib_visc_calc2,bib_visc_calc3}. 

 The $v_n$ can be measured by the event plane (EP) method where one determines the orientation of the harmonic planes $\Psi_n$ and then, by measuring the yield of the particles about these planes, obtains the $v_n$. They can also be measured by the two particle correlation (2PC) method where one measures the relative yield of associated (or partner) particles with respect to trigger particles in $\Delta\phi=\phi_{trigger}-\phi_{partner}$ and $\Delta\eta=\eta_{trigger}-\eta_{partner}$. The 2PC correlation function can be expanded in a Fourier series in $\Delta\phi$ as: 

\begin{equation}
C(\Delta\phi) \propto 1+2\sum_{n=1}^{\infty} v_{n,n}(p_T^a,p_T^b) \cos(n\Delta\phi)
\label{eq_correlation}
\end{equation} 

Where the superscripts $a$ and $b$ label the trigger and partner particles. If the only contribution to correlations comes from collective flow, then the Fourier coefficients $v_{n,n}$ are equal to the product of the individual single particle $v_n$ \cite{bib_factorization}:

\begin{equation}
v_{n,n}(p_T^a,p_T^b)=v_n(p_T^a)\times v_n(p_T^b)
\label{eq_scaling_relation}
\end{equation} 

Using Eq.\ref{eq_scaling_relation}, one can obtain the $v_n$ from the 2PC. The above relation is violated if non-flow effects, like jets, are present. Thus before extracting the flow harmonics from the correlations, one must ensure that most (if not all) of the non-flow effects are removed.

All the results presented in this proceeding are for inclusive charged hadrons reconstructed using the ATLAS \cite{atlas_detector} inner detector covering the pseudorapidity range $|\eta|<2.5$. For the event plane analysis, the harmonic planes are determined using the Q-vector method \cite{bib_q_vector}. The Q-vectors are measured using the ATLAS forward calorimeter (FCal) which lies at $|\eta|\in(3.3,4.8)$. Details of the results presented here can be found in \cite{bib_vn_note,atlas_flow_paper}.

\section{Event plane analysis}


The measured $\Psi_n$ vary around the true $\Psi_n$ due to statistical fluctuations \cite{bib_q_vector}. In order to correct for this, the $v_n$ are obtained by first calculating $v_n^{obs}=\langle \cos n(\phi-\Psi_n)\rangle$ and then dividing the $v_n^{obs}$ by the reaction plane resolution \cite{bib_q_vector}.

\begin{equation}
v_n=v_n^{obs}/Resolution(\Psi_n)
\end{equation}

\begin{figure}[ht!]
\centering
\includegraphics[width=80mm]{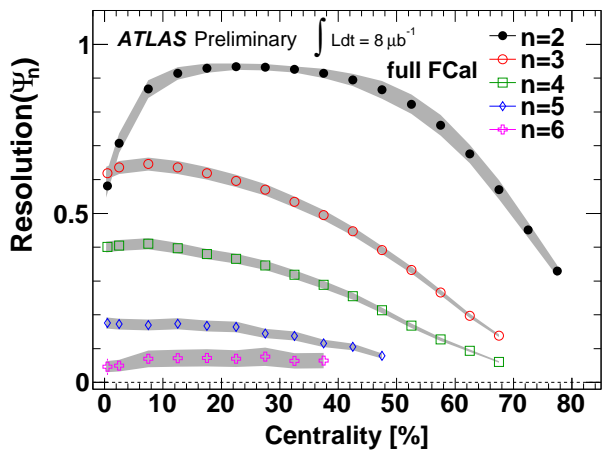}
\caption{FCal reaction plane resolutions for harmonics 2 to 6. Bands indicate systematic uncertainties.} \label{fig_rpreso}
\end{figure}

The FCal reaction plane resolutions for $\Psi_n$, $n$=2-6 are shown in Fig.\ref{fig_rpreso}. For different harmonics, the resolutions are cut off at different centralities depending on where they become small as compared to the systematic and statistical uncertainties.

\begin{figure}[ht!]
\centering
\includegraphics[width=110mm]{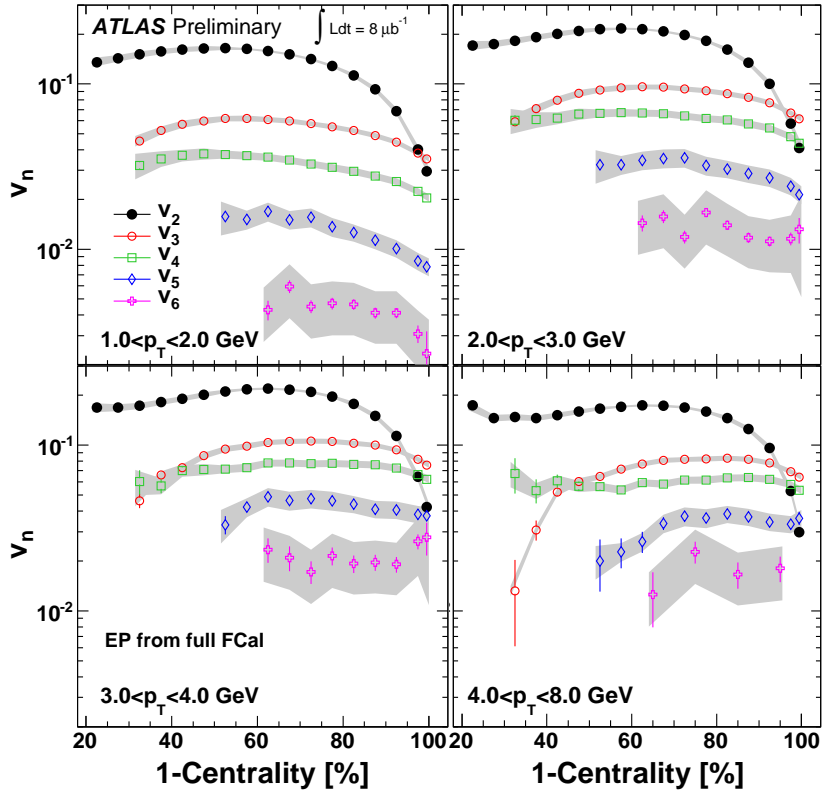}
\caption{$v_n$ as a function of centrality is 5\% centrality bins plus a 0-1\% most central bin (the right most point) for four $p_T$ bins. The shaded bands indicate systematic uncertainties.} 
\label{fig_vncent_rp}
\end{figure}

In Fig.\ref{fig_vncent_rp} the $v_n$ are plotted as a function of centrality from EP method in 5\% centrality bins plus a 0-1\% super-central bin. Each panel is a different $p_T$ range. It is seen that $v_2$ has a much stronger centrality dependence as compared to the other harmonics.  For non-central events, $v_2$ becomes much larger than the other $v_n$ due to the elliptical geometry of the fireball. The other $v_n$, mainly driven by event-by-event fluctuations in the initial geometry have a weaker centrality dependence. 
This centrality dependence of the collision geometry is also the reason why the reaction plane resolution for $\Psi_2$ shown in Fig.\ref{fig_rpreso} increases from central to mid central events, while the resolution for the other $\Psi_n$ do not show such a strong centrality dependence.
Also note that in most central events (top 5\% or top 1\%) and at sufficiently high $p_T$,  $v_3$, $v_4$ and even $v_5$ can become larger than or comparable to $v_2$.

\begin{figure}[ht!]
\centering
\includegraphics[width=135mm]{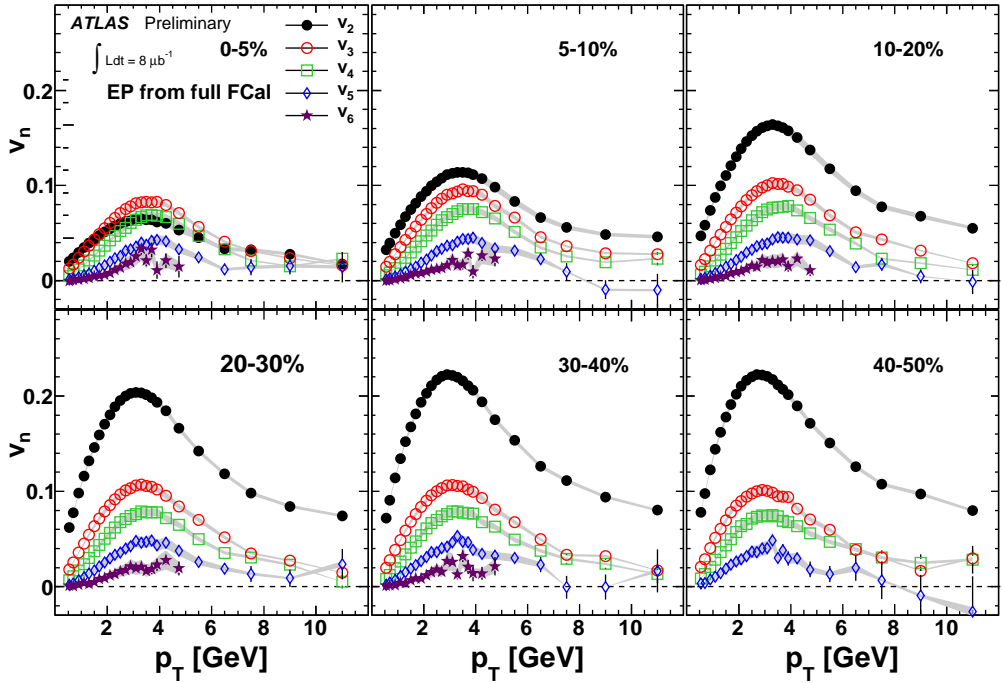}
\caption{$v_n$ vs. $p_T$ for several centrality classes. The shaded bands indicate the systematic uncertainties. Note that the $v_6$ is only measured for 0-40\%.}
\label{fig_vnpt_rp}
\end{figure}

\begin{figure}[ht!]
\centering
\includegraphics[width=135mm]{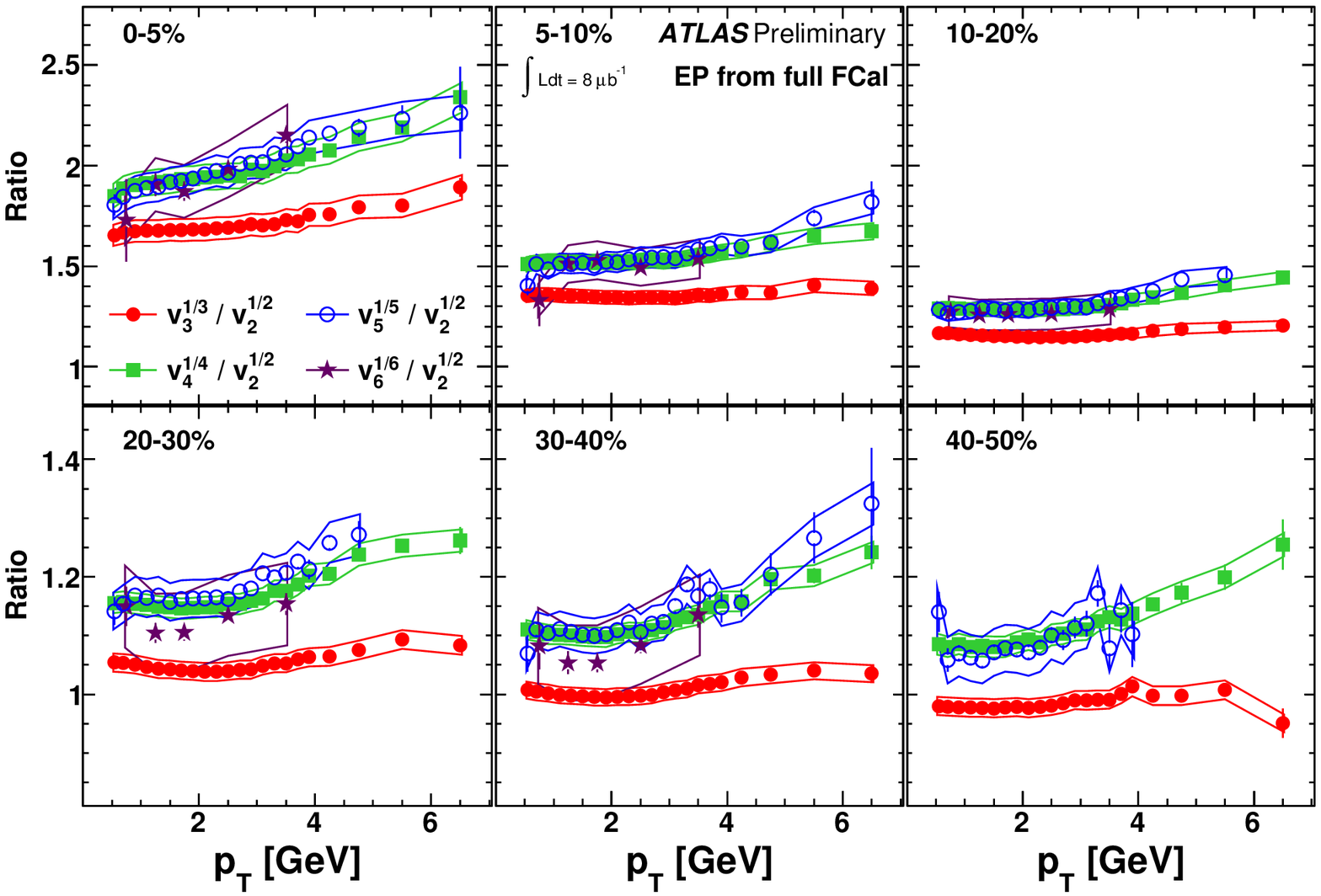}
\caption{$v_n^{1/n}/v_2^{1/2}$ vs. $p_T$ for several centrality classes. Lines indicate systematic uncertainty bands, calculated by assuming that the uncertainties for different $v_n$ are independent.}
\label{fig_vnptscale_rp}
\end{figure}

Figure \ref{fig_vnpt_rp} shows $v_n$ as a function of $p_T$. For all centralities the $v_n$ have a similar trend, they increase to reach a maximum between 3-4 GeV and then decrease. In the 0-5\% central events, $v_3$ and $v_4$ become larger than $v_2$ at intermediate $p_T$. A scaling, previously observed at RHIC \cite{bib_vn_scaling} is seen in the $p_T$ dependence of $v_n$: $v_n^{1/n}=k_n v_2^{1/2}$ where $k_n$ are only weakly dependent on $p_T$. In Fig.\ref{fig_vnptscale_rp} the values of $k_n$ are plotted as a function of $p_T$ for different centralities. The $k_n$, though weakly dependent on $p_T$, have a strong centrality dependence.

\begin{figure}[ht!]
\centering
\includegraphics[width=135mm]{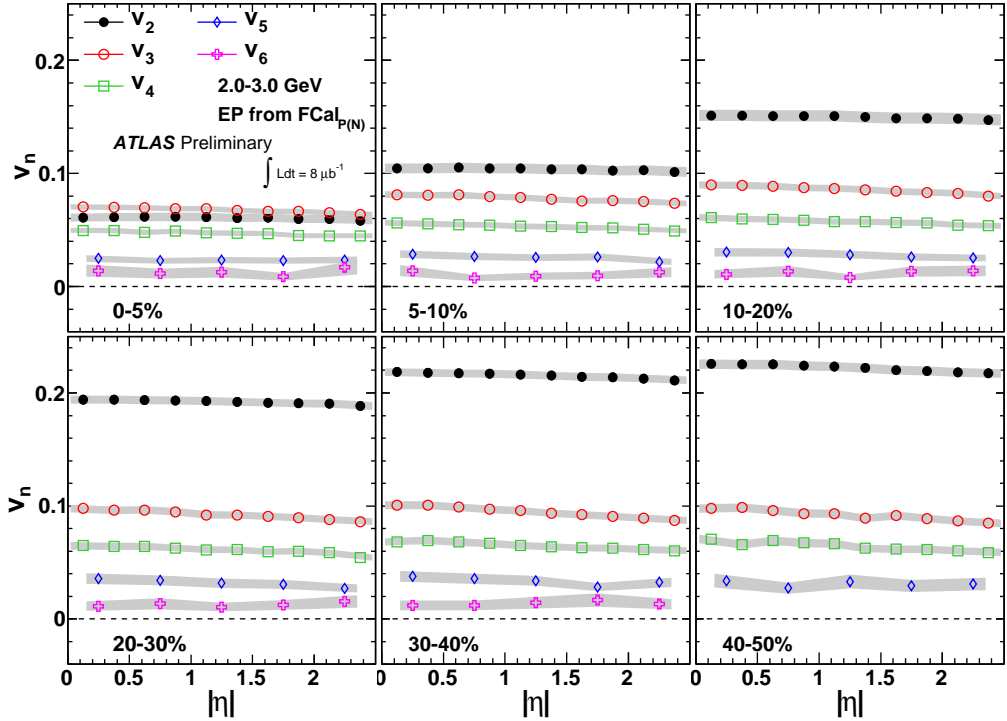}
\caption{$\eta$ dependence of $v_n$ for $p_T\in(2,3)$GeV. Each panel is a different centrality bin.} 
\label{fig_vneta_rp}
\end{figure}

\begin{figure}[ht!]
\centering
\includegraphics[width=135mm]{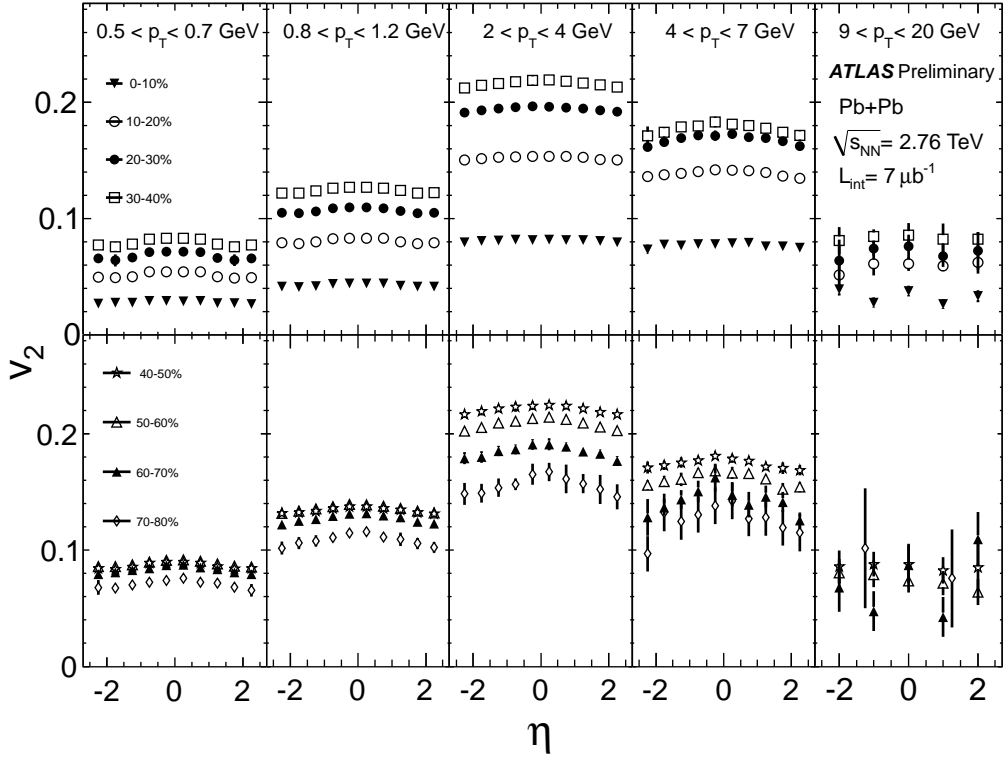}
\caption{$\eta$ dependence of $v_2$ for $0.5< p_T < 20$ GeV in five $p_T$ intervals and 10\% centrality intervals. Error bars show statistical and systematic uncertainties added in quadrature.}
\label{fig_vneta_rp2}
\end{figure}

In Fig.\ref{fig_vneta_rp} the $\eta$ dependence of the $v_n$ are shown for $2<p_T<3$ GeV. In this $p_T$ interval, $v_2$ decreases by at most 5\% from $\eta=0$ to $\eta=2.5$, for $n>2$ the decrease is slightly larger. In general, there is a weak dependence of the differential $v_n$ on $\eta$ in central and mid-central collisions. For $v_2$, this weak $\eta$ dependence is demonstrated over a large $p_T$ and centrality range in Fig.\ref{fig_vneta_rp2}. 

Eq.\ref{eq_scaling_relation} stated that for pure flow, the 2PC $v_{n,n}$ is the product of the single particle $v_n$. This relation is only true if there is no $\eta$ dependence in the $v_n$ (or approximately true if the $\eta$ dependence is weak). Otherwise Eq.\ref{eq_scaling_relation} would have to be replaced with a convoluted integral of the $\eta$ dependent $v_n$ weighted by the $\eta$ dependent spectra. The weak $\eta$ dependence justifies the use of Eq.\ref{eq_scaling_relation}.

\section{Two particle correlations}
\begin{figure}[ht!]
\centering
\includegraphics[width=135mm]{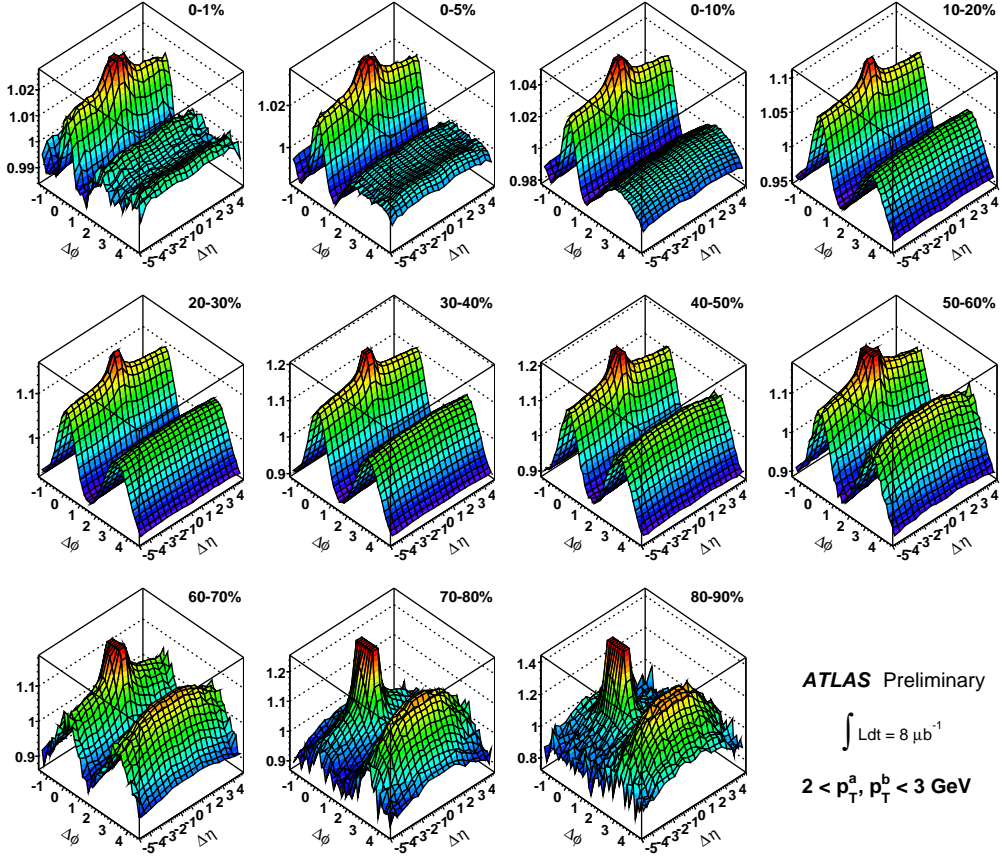}
\caption{Two particle $\Delta\eta-\Delta\phi$ correlations for $2<p_T^a,p_T^b<3$GeV. Each panel is a different centrality bin. The peripheral events have the near side jet peak truncated.} 
\label{fig_2pc}
\end{figure}

Figure \ref{fig_2pc} shows the 2PC in $\Delta\eta$-$\Delta\phi$ space measured in the momentum range $2<p_T^a,p_T^b<3$ GeV for several centrality classes. The near side jet peak reported in \cite{bib_ridge_cone1} is clearly visible at $(\Delta\eta,\Delta\phi)\sim(0,0)$ across all centralities and is more prominent in peripheral events.
The near side ridge has a weak $\Delta\eta$ dependence in central and mid-central collisions. The strength of the ridge increases from central to mid-central events and then decreases for peripheral events, completely disappearing in the 80-90\% centrality class, leaving behind only the near-side jet peak. In the most central events (first and second panels) one can see a double hump structure on the away side which transforms into a ridge-like structure in the mid-central events. In peripheral events, it is replaced by the away side jet.

The presence of the near side jet biases the values of the flow harmonics obtained from the two particle correlations. This bias can be minimized by requiring a $|\Delta\eta|>2.0$ cut between the trigger and the associated particle in the correlations. While this cut removes the near side jet, part of the away side jet is still left. This is because the away side jet is not localized in $\Delta\eta$ and cannot be removed by a cut on $|\Delta\eta|$. Thus the measured $v_{n,n}$ still retain non-flow bias.

\begin{figure}[ht!]
\centering
\includegraphics[width=110mm]{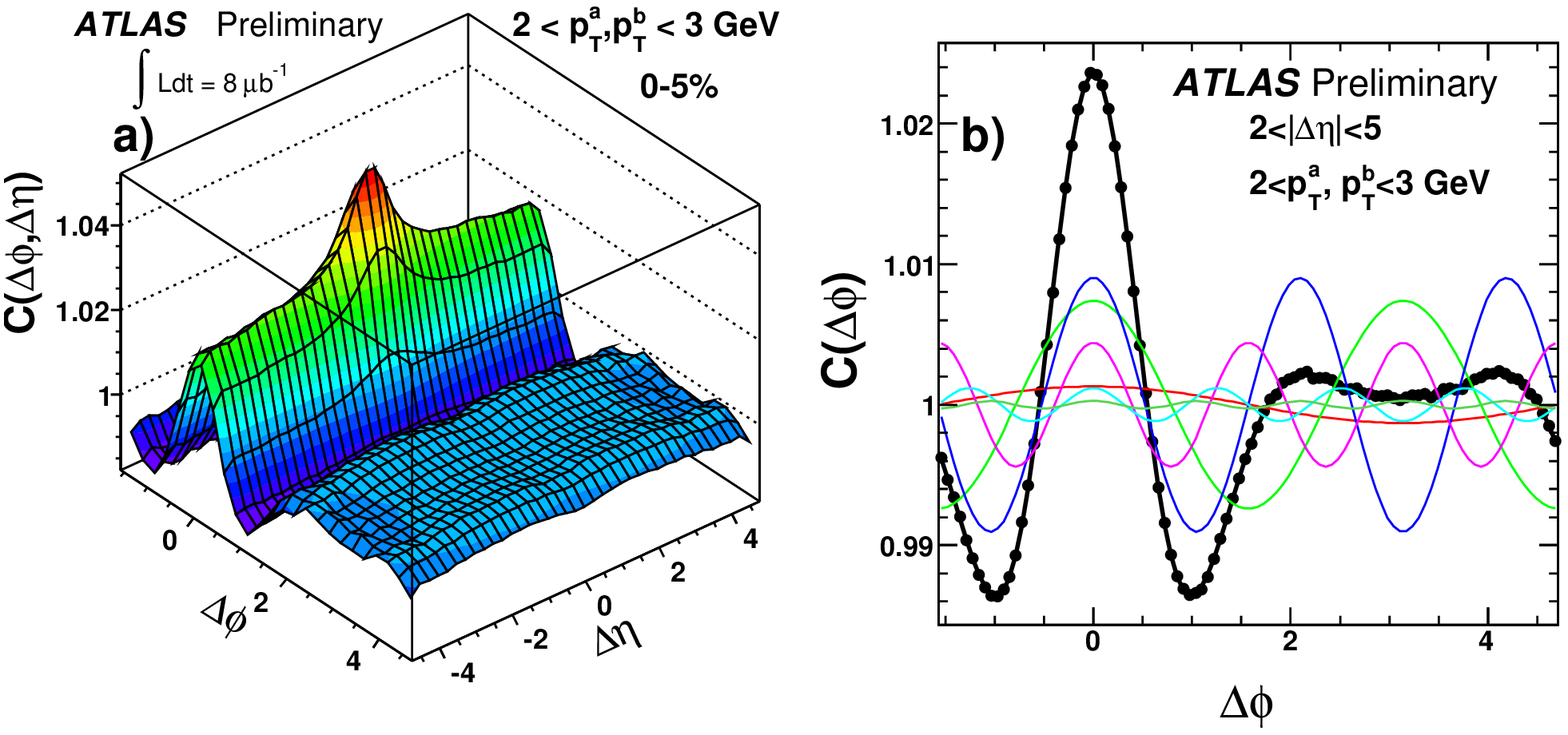}
\includegraphics[width=110mm]{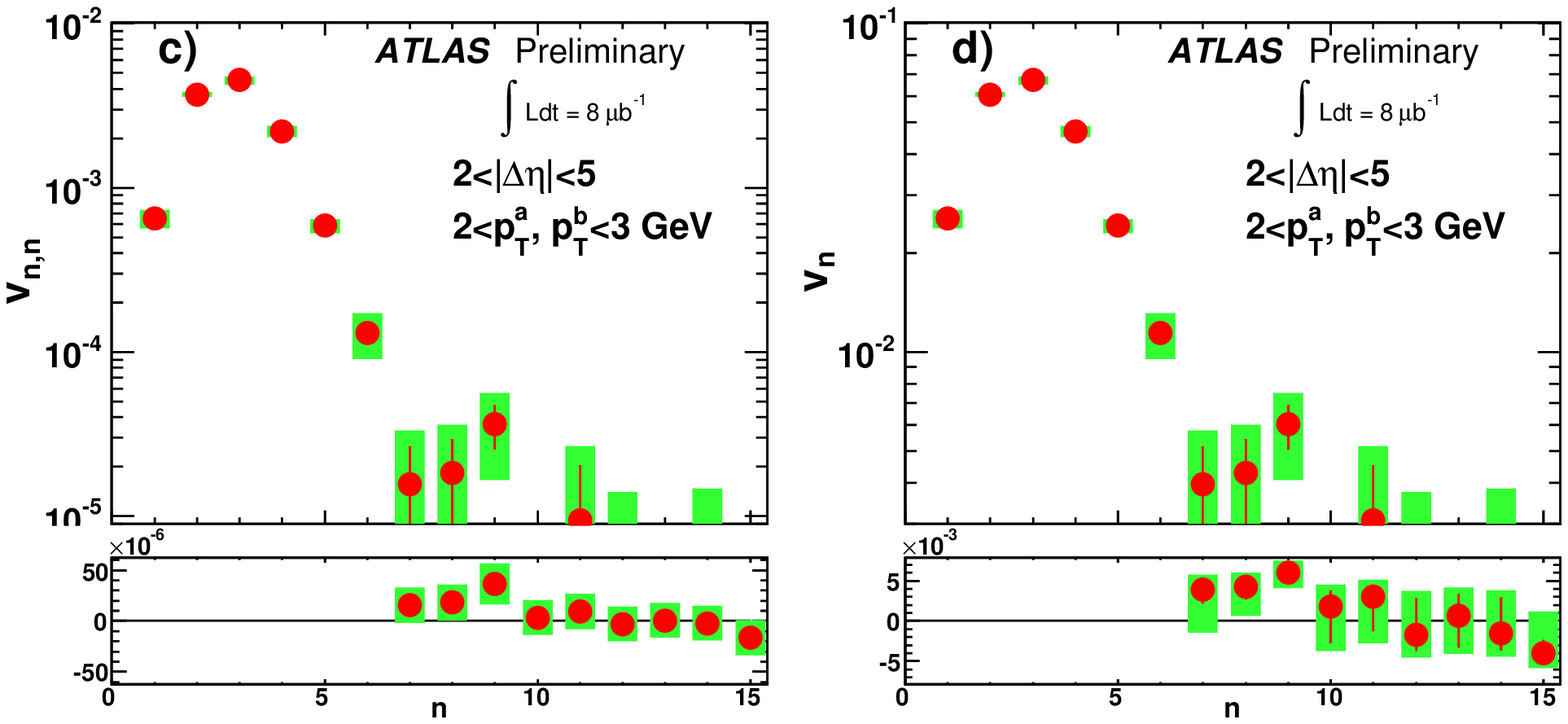}
\caption{The steps involved in the extraction of the $v_n$ for 2-3 GeV fixed-$p_T$ correlation. Panel a): the two-dimentional correlation function (shown for $|\Delta\eta|<4.75$ to reduce fluctuations near the edge). Panel b): The one-dimentional $\Delta\phi$ correlation function for $2<|\Delta\eta|<5$. Panel c): $v_{n,n}$ vs. $n$. Panel d): $v_{n}$ vs. $n$. The lower sub panels in c) and d) also show the $v_{n,n}$ and $v_n$ respectively for $n\geq7$ but in a linear scale.}
\label{fig_2pc_procedure}
\end{figure}

Figure \ref{fig_2pc_procedure} explains the procedure used for obtaining the $v_n$ from the 2PC. The panel a) shows the two-dimensional correlation function in $\Delta\eta$-$\Delta\phi$ for $2<p_T^a,p_T^b<3$ GeV. Such correlations, where both trigger and partner have same $p_T$ range, are termed as \textit{fixed}-$p_T$ correlations. The panel b) shows the the one-dimensional $\Delta\phi$ correlation function (black points) in $\Delta\phi$ for $2<|\Delta\eta|<5$, overlaid with contributions from individual Fourier components and their sum. As mentioned before, the $|\Delta\eta|>2.0$ cut is imposed to remove the near side jet. The Fourier coefficients $v_{n,n}$ obtained from the one-dimensional correlation are shown in panel c) as a function of $n$. The bars and bands indicate statistical and systematic uncertainties respectively. As the trigger and partner $p_T$ ranges are chosen to be the same, Eq.\ref{eq_scaling_relation} becomes:

\begin{equation}
v_{n,n}(p_T^a,p_T^a)=v_n^2(p_T^a) 
\label{eq_scaling_relation2}
\end{equation} 

Using the above relation one calculates the values of $v_n$ and its uncertainties from $v_{n,n}$. The $v_n$ values are plotted as a function of $n$ in the panel d) of Fig.\ref{fig_2pc_procedure}. In panels c) and d) of Fig.\ref{fig_2pc_procedure} the $v_{n,n}$ and $v_n$ values are plotted untill $n=15$. However, the analysis is limited to $n\leq6$ because for higher $n$ the systematic and statistical uncertainties are large compared to the extracted value of $v_{n,n}$.

\begin{figure}[ht!]
\centering
\includegraphics[width=135mm]{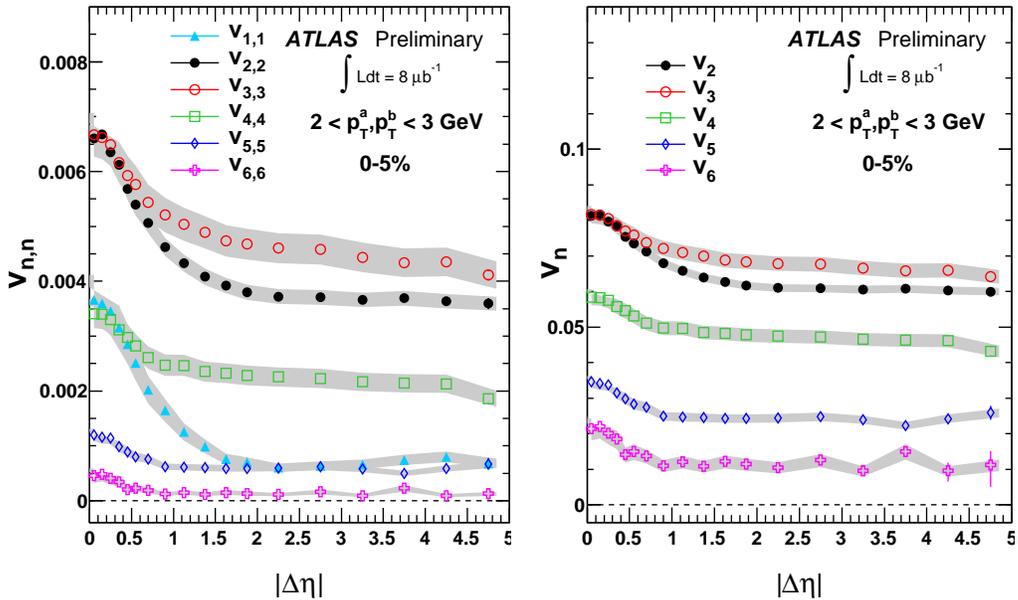}
\caption{$v_{n,n}$ and $v_n$ as a function of $\Delta\eta$ for 2-3 GeV fixed-$p_T$ correlation for events in 0-5\% centrality class.} 
\label{fig_2pc_etacut_explanation}
\end{figure}

As mentioned before, a $|\Delta\eta|>2.0$ cut was used to remove the effects of the near side jet. The choice of this particular value of the cut can be justified in the following way. Figure \ref{fig_2pc_etacut_explanation} shows the $v_{n,n}$ and $v_n$ obtained for 2-3 GeV fixed-$p_T$ correlation as a function of $|\Delta\eta|$. The $v_{nn}$ and $v_n$ show a clear peak at $|\Delta\eta|\sim0$ due to the near side jet. For larger values of $|\Delta\eta|$, the $v_{n,n}$ values drop as the influence of the near side jet weakens. For $|\Delta\eta|>2$, the fractional change is very small.

The two-particle correlation analysis used the fixed-$p_T$ correlations and Eq.\ref{eq_scaling_relation2} to obtain the $v_n$ values from the $v_{n,n}$. The $v_n$ values obtained from fixed-$p_T$ can be cross-checked using correlations where the trigger and partner have different $p_T^a$ and $p_T^b$ values. Such correlations are termed as \textit{mixed}-$p_T$ correlations. The $v_n(p_T^b)$ of the partner can be obtained as:

\begin{equation}
v_n(p_T^b)=v_{n,n}(p_T^a,p_T^b)/v_n(p_T^a)
\label{eq_mixed_relation}
\end{equation}

\begin{figure}[ht!]
\centering
\includegraphics[width=135mm]{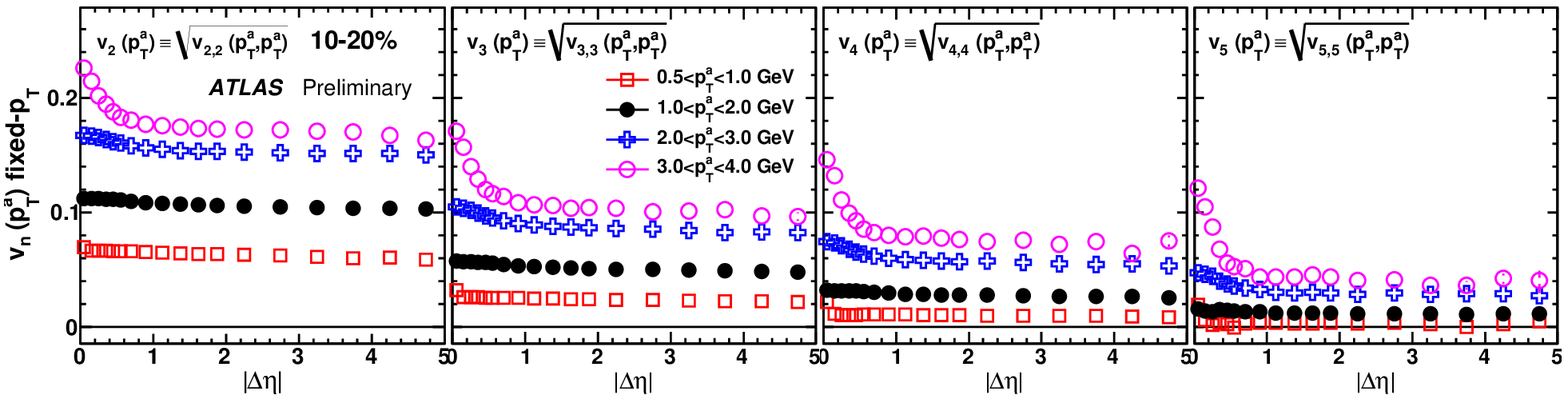}
\includegraphics[width=135mm]{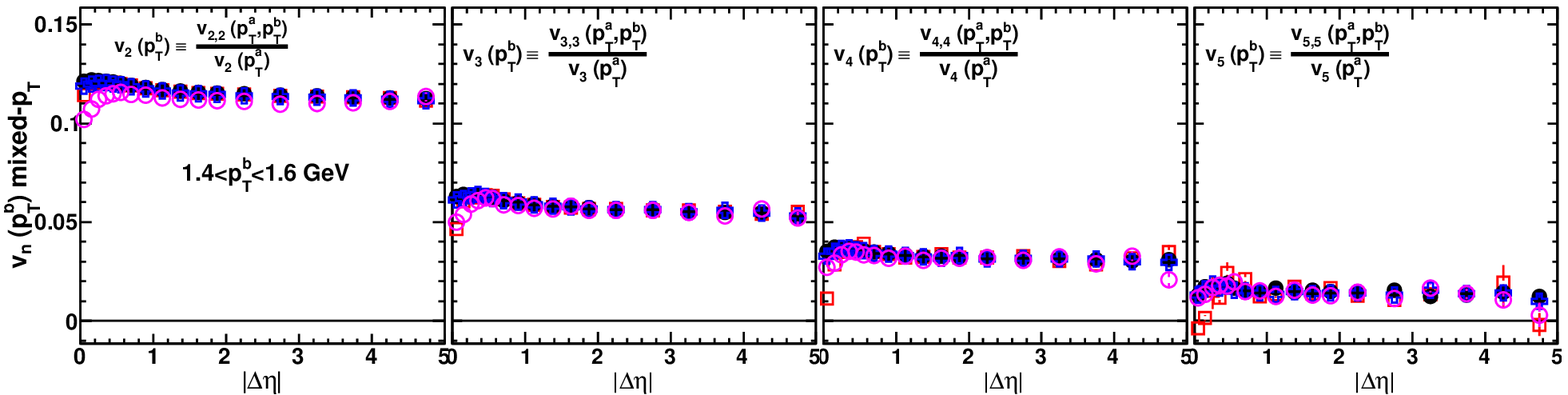}
\caption{Top panels: $v_n(p_T^a)=\sqrt{|v_{n,n}(p_T^a,p_T^a)|}$ ($n=2-5$ from left to right) vs. $|\Delta\eta|$ for four fixed-$p_T$ correlations ($p_T^a\in$ 0.5-1, 1-2, 2-3 and 3-4 GeV ranges). Bottom panels: $v_n(p_T^b)=v_{n,n}(p_T^a,p_T^b)/v_{n}(p_T^a)$ vs. $|\Delta\eta|$ for $p_T^b\in(1.4,1.6)$GeV.} 
\label{fig_2pc_universality}
\end{figure}

 Equation \ref{eq_mixed_relation} can be checked by measuring the same $v_n(p_T^b)$ for different $v_n(p_T^a)$. This is illustrated in Fig.\ref{fig_2pc_universality}. In the top panels the $v_2-v_5$ for four different $p_T$ obtained from fixed-$p_T$ correlations bins are plotted as a function of $|\Delta\eta|$ for the 10-20\% centrality events. For these four values of the trigger $p_T$, mixed-$p_T$ correlations are done with the associated particle $p_T^b\in(1.4.1.6)$ GeV. The bottom panels of Fig.\ref{fig_2pc_universality} show the $v_n(p_T^b)$ obtained from the mixed-$p_T$ correlations using Eq.\ref{eq_mixed_relation}. At low $|\Delta\eta|$ different triggers give different values for $v_n(p_T^b)$ showing that the scaling relation Eq.\ref{eq_scaling_relation} (or Eq.\ref{eq_mixed_relation}) breaks down due to the effects of the near side jet. For $|\Delta\eta|>1.0$ however, all trigger-partner combinations give nearly identical values of $v_n(p_T^b)$ even though the trigger $v_n$ values vary over a large range. This validates Eq.\ref{eq_scaling_relation} and Eq.\ref{eq_mixed_relation}, and shows that indeed the Fourier components in the 2PCs at large $|\Delta\eta|$ are due to collective flow. Note that the mixed-$p_T$ correlations are only used as a cross-check to qualitatively check that Eq.\ref{eq_scaling_relation} works, they are not used to determine any systematic errors for the 2PC $v_n$.

\section{Comparison of $v_n$ obtained from 2PC and EP methods}

\begin{figure}[ht!]
\centering
\includegraphics[width=110mm]{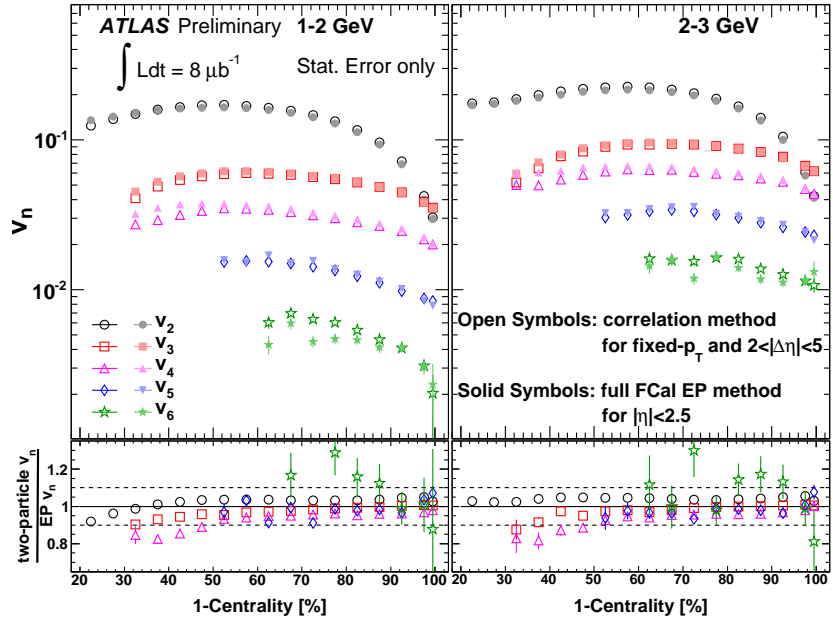}
\caption{Comparison of $v_n$ obtained from the two-particle fixed-$p_T$ correlation method (solid symbols) to the EP results (open symbols). The ratios between the two methods are shown in the bottom panels. Results for two $p_T$ bins are shown: $1-2GeV$ (left panels) and $2-3GeV$ (right panels). The dashed lines indicate a $\pm10$\% band to guide the eye.} 
\label{fig_compare_centdep}
\end{figure}

Figure \ref{fig_compare_centdep} shows the centrality dependence of the $v_n$ obtained from the two methods. For clarity only the statistical errors are shown. From the ratio plots in the bottom panels one can see that the values for $v_2$, $v_3$ and $v_4$ agree within 5\% for central and mid central events. For $v_5$ and $v_6$ they agree within 10\% and 15\% respectively and are consistent within the systematic uncertainties quoted in \cite{bib_vn_note}.

\begin{figure}[ht!]
\centering
\includegraphics[width=135mm]{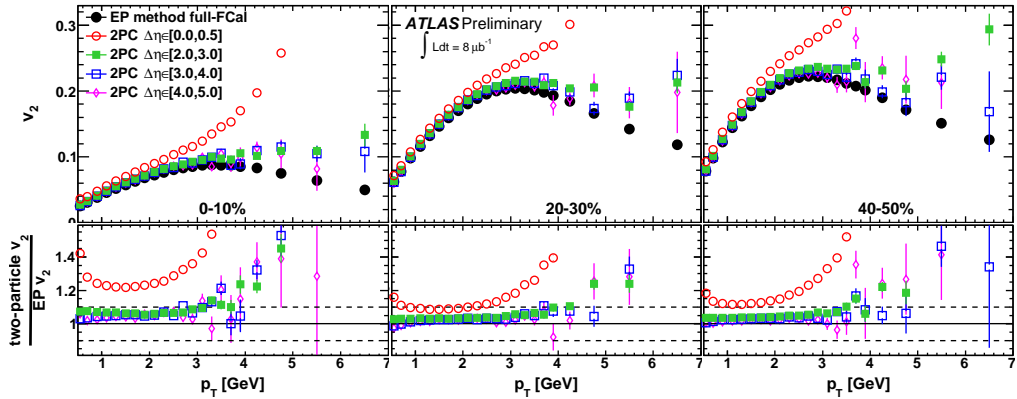}
\caption{Top panels: Comparison of $v_2$ determined using the EP method and the two-particle fixed-$p_T$ correlation method in several $\Delta\eta$ slices for three centrality bins. Bottom panels: Ratios of $v_2$ between the two-particle and EP results.}
\label{fig_compare_ptdep}
\end{figure}

In Fig.\ref{fig_compare_ptdep} the $p_T$ dependence of $v_2$ obtained from two-particle fixed-$p_T$ correlation method in several $\Delta\eta$ slices are compared to the EP values. 

The 2PC values measured for $|\Delta\eta|\in(0,0.5)$ for all $p_T$ are signiﬁcantly higher than the EP results due to the bias introduced by the near side jet. However, the curves measured with $|\Delta\eta|>2.0$ agree much better below $p_T$=4.0 GeV. For higher values the away side jet presumably affects the 2PC results which explains their deviation from the EP $v_n$ even for $|\Delta\eta|>2.0$.

For central and mid central events and $p_T<4$ GeV, the two particle correlation with $|\Delta\eta|>2$ gap and EP method give consistent values of $v_n$. This indicates that the structures in two particle correlations at large $\Delta\eta$ including the ridge and cone can be entirely accounted for by collective flow. To demonstrate this clearly, the EP $v_n$ values are used to reconstruct the 2PC which are then compared to the measured 2PC. The reconstructed $v_{n,n}$ are calculated according to Eq.\ref{eq_scaling_relation} as the product of the $v_n$ obtained using the EP method:

\begin{equation}
v_{n,n}^{reco}=v_n^{EP}\times v_n^{EP}
\label{eq_reco_vnn}
\end{equation}

Since $v_1$ measurements were not made using the EP method, the $v_{1,1}$ component of the reconstructed correlation function is taken to be the same as that of the measured correlation function. The overall normalization of the reconstructed correlation function is taken to be the same as that of the measured correlation function ($N_0^{2PC}$). The final expression for the reconstructed correlation function can be written as:

\begin{equation}
C^{reco}(\Delta\phi)=N_0^{2PC} \biggl( 1+2v^{2PC}_{1,1}\cos(\Delta\phi)+ 2\sum_{n=2}^{6}v_n^{EP}v_n^{EP}\cos(n\Delta\phi) \biggr)
\label{eq_reco_corr}
\end{equation}

Where, as explained before the normalization and the $v_{1,1}$ component are taken to be the same as that of the measured correlation.

\begin{figure}[Ht!]
\centering
\includegraphics[width=135mm]{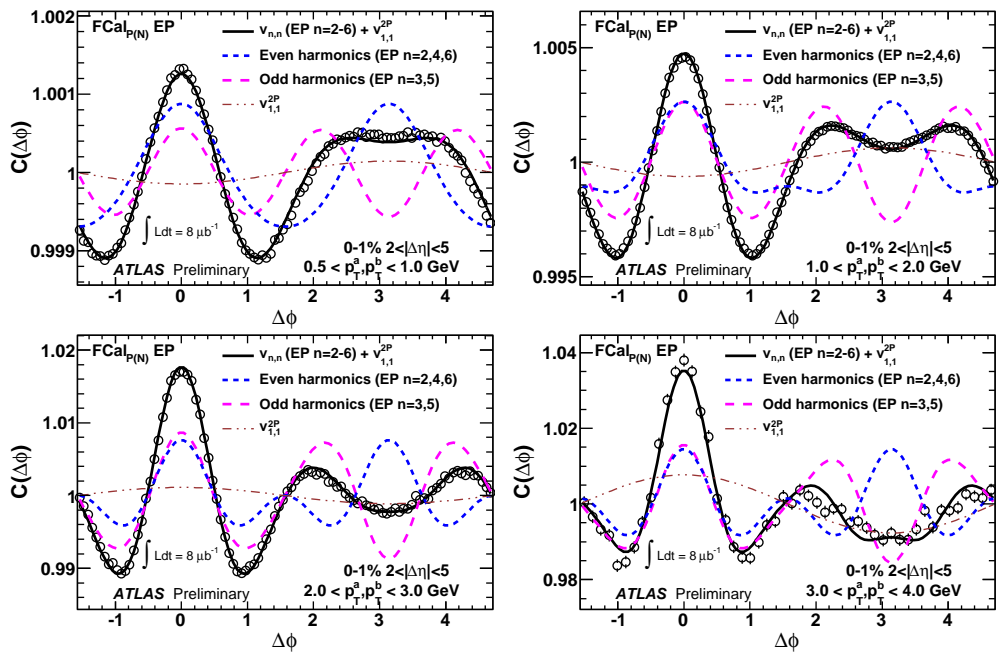}
\caption{Measured correlation function data (circles) compared with that reconstructed from $v_{1,1}$ from two-particle correlations and $v_2-v_6$ from EP method for 0-1\% centrality for several fixed-$p_T$ correlations.}
\label{fig_reconstruct_correlation}
\end{figure}

In Fig.\ref{fig_reconstruct_correlation} four reconstructed correlation functions are compared to the corresponding measured  ones. The 2PC are well reproduced with both the near side ridge as well as the away side double hump. 

\section{Summary}



To summarize, the flow harmonics $v_2$-$v_6$ are measured by event plane method as well as two particle correlation method. These measurements were done to higher orders, n=6 and over a much larger $p_T$ and $\eta$ ranges than previously done at RHIC. An approximate scaling relationship between the $v_n$: $v_n^{1/n}\propto v_2^{1/2}$ which was observed at RHIC was shown to hold also at LHC energies. 

For $|\Delta\eta|\sim0$ the $v_n$ values obtained from 2PC are significantly different than the EP values, due to the effects of the near side jet. However, as the $\eta$ gap between the trigger and partner is increased, the 2PC $v_n$ values become consistent with the EP ones for central and mid-central events.

It was demonstrated that the ridge and cone seen in two particle correlations at low and intermediate $p_T$ ($p_T<4.0$ GeV) can be entirely accounted for by the collective flow of the medium. There is no need for any additional physics models to account for these structures.

\begin{acknowledgments}
This work is in part supported by NSF under award number PHY-1019387.
\end{acknowledgments}

\bigskip 

\begin{thebibliography}{9}   

\bibitem{bib_q_vector}  A. M. Poskanzer and S. A. Voloshin, Phys. Rev. C 58, 1671 (1998).
\bibitem{bib_ridge_cone1} PHENIX Collaboration, A. Adare \textit{et al}., Phys. Rev. C 78, 014901 (2008).
\bibitem{bib_ridge_cone2} STAR Collaboration, B. I. Abelev \textit{et al}. Phys. Rev. C 80, 064912 (2009).
\bibitem{bib_v3cone_} B. Alver and G. Roland, Phys. Rev. C 81, 054905 (2010) [Erratum-ibid. C 82, 039903 (2010)]
\bibitem{bib_visc_calc1} Z. Qiu and U. W. Heinz, arXiv:1104.0650 [nucl-th].
\bibitem{bib_visc_calc2} B. Schenke, S. Jeon, C. Gale, arXiv:1102.0575 [hep-ph].
\bibitem{bib_visc_calc3} G. Y. Qin, H. Petersen, S. A. Bass and B. Muller, Phys. Rev. C 82, 064903 (2010).
\bibitem{bib_factorization} R. A. Lacey, Nucl. Phys. A698, 559 (2002); R. J. M. Snellings, \textit{ibid.} A698, 193 (2002).

\bibitem{atlas_detector} ATLAS Collaboration, JINST 3, S08003 (2008).
\bibitem{bib_vn_note} ATLAS Collaboration, Measurement of elliptic flow and higher-order flow coefficients with the ATLAS detector in $\sqrt{s_{NN}}$=2.76 TeV Pb+Pb collisions, \url{http://cdsweb.cern.ch/record/1349924}.
\bibitem{atlas_flow_paper} ATLAS Collaboration, arXiv:1108.6018 [nucl-ex].
\bibitem{bib_vn_scaling} R. Lacey \textit{et al}, arXiv:1105.3782 [nucl-ex].

\end{thebibliography}

\end{document}